\documentclass{acm_sen_article}

\usepackage[numbers]{natbib} 

\usepackage{lipsum}
\usepackage{url}
\usepackage{flushend}
\usepackage{xcolor}

\usepackage{lipsum}
\usepackage{url}
\usepackage{flushend}
\usepackage{xcolor}

\begin{document}

\title{Blended PC Peer Review Model: Process and Reflection}

\numberofauthors{5} 
\author{
Chakkrit Tantithamthavorn\\
       \affaddr{Monash University}\\
       \affaddr{Melbourne, Australia}\\
       \email{chakkrit@monash.edu}
\and
Nicole Novielli\\
       \affaddr{University of Bari}\\
       \affaddr{Bari, Italy}\\
       \email{nicole.novielli@uniba.it}
\and 
Ayushi Rastogi\\
       \affaddr{University of Groningen}\\
       \affaddr{Groningen, Netherlands}\\
       \email{a.rastogi@rug.nl}
 \and 
Olga Baysal\\
       \affaddr{Carleton University}\\
       \affaddr{Ottawa, Canada}\\
       \email{olga.baysal@carleton.ca}
\and 
Bram Adams\\
       \affaddr{Queen's University}\\
       \affaddr{Kingston, Canada}\\
       \email{bram.adams@queensu.ca}       
}

\maketitle







\begin{abstract}
The academic peer review system is under increasing pressure due to a growing volume of submissions and a limited pool of available reviewers, resulting in delayed decisions and an uneven distribution of reviewing responsibilities. 
Building upon the International Conference on Mining Software Repositories (MSR) community's earlier experience with a Shadow PC (2021 and 2022) and Junior PC (2023 and 2024), MSR 2025 experimented with a Blended Program Committee (PC) peer review model for its Technical Track. This new model pairs up one Junior PC member with two regular PC members as part of the core review team of a given paper, instead of adding them as an extra reviewer. This paper presents the rationale, implementation, and reflections on the model, including empirical insights from a post-review author survey evaluating the quality and usefulness of reviews. Our findings highlight the potential of a Blended PC to alleviate reviewer shortages, foster inclusivity, and sustain a high-quality peer review process. We offer lessons learned and recommendations to guide future adoption and refinement of the model.
\end{abstract}

\maketitle

\section{Introduction and Background}
The academic peer review system is facing increasing strain due to a shortage of qualified reviewers and the resource-intensive nature of the process~\cite{petrescu2022evolving}. 
As the volume of paper submissions continues to rise, 
software engineering conferences, including the International Conference on Mining Software Repositories (MSR),
struggle to secure timely and high-quality reviews~\cite{bacchelli2017double}. 

Many experienced researchers decline review invitations due to overwhelming workloads, resulting in delays and an uneven distribution of the reviewing burden, while early-career researchers often struggle to receive invitations despite their willingness and capability to contribute.
This shortage not only compromises the efficiency of the academic peer review system but also raises concerns about the diversity, equity, and inclusivity, and long-term sustainability of the peer review ecosystem.

In 2005, SIGCOMM --- an annual flagship conference on Data Communications introduced the Shadow PC initiative. 
Inspired by a similar initiative at NSDI\footnote{https://www.usenix.org/conference/nsdi15/call-for-papers}, the goal of the Shadow PC initiative was to counter biased program committee demographics and improve transparency~\cite{feldmann2005experiences}.
This pilot mirrored the main PC process, including reviews and meetings, enabling those outside the core community to gain insight into the review process and understand quality through exposure to both strong and weak submissions~\cite{feldmann2005experiences}. 

Later in 2007, the SOSP conference adopted the model~\cite{isaacs2008report}, shifting the focus to education. 
Their Shadow PC program welcomed junior researchers globally and aimed to build reviewing skills, confidence, and long-term engagement. 
Unlike SIGCOMM's approach, SOSP emphasized mentorship and reflective learning by comparing Shadow and regular PC review decisions~\cite{isaacs2008report}. 
More recently (2024), the HCI community ran a five-month program focused on equity and inclusion, involving early-career researchers from 26 countries, many of whom were new to reviewing, including under-resourced settings in Africa and Asia~\cite{varghese2024shadow}.

Recently, conferences like AAAI have experienced a reviewer shortage despite lowering the bar to recruit reviewers. Thereafter, in 2021, the ICML conference proposed a novel approach to train less-experienced researchers to contribute alongside seasoned reviewers~\cite{stelmakh2021novice}. 
The study recruited less-experienced researchers through a review test, following which their participation was evaluated through both indirect indicators (e.g., review length, discussion activity) and meta-reviewer assessments. 
The study shows that their reviews were comparable in quality to the main PC. 

In software engineering, the International Conference on Mining Software Repositories (MSR) introduced the Shadow PC initiative in 2021. 
Like its predecessors in other fields, participants in the Shadow PC initiative were early-career researchers involved in a parallel review process for training purposes without a direct influence on the outcome of the main track reviewing.

An early-career researcher is a PhD student, postdoc, new faculty member, or industry practitioner who is keen to get more involved in the academic peer-review process but has not yet served on a technical research track program committee at respected international software engineering conferences (e.g., ICSE, ESEC/FSE, ASE, MSR, ICSME, SANER).
Early career researchers were also offered training by distinguished reviewers in the field to write quality reviews. 
This approach showed consistent effectiveness of training early career researchers~\cite{thongtanunam2021shadow}. 

Inspired by the success of the MSR Shadow PC initiative in 2021 and 2022, in 2023, the Junior PC initiative was launched. 
The ultimate goal of the Junior PC initiative was to train early-career researchers in a real environment.
Unlike the Shadow PC initiative, in the Junior PC initiative, participants were involved in the actual review process and contributed to the final decision of the papers submitted, of course, with mentoring support. 
So, every paper was reviewed by two junior PC members in addition to 3 main PC members. 
While effective, it raised practical challenges, especially in reconciling multiple viewpoints and complicating author responses.

This paper presents the Blended Program Committee (PC) peer review model proposed and implemented by the 2025 International Conference on Mining Software Repositories (MSR), held in Ottawa, Canada. 
Building on previous initiatives within and outside software engineering, the goal is to keep the process lightweight and sustainable. 

To keep the process lightweight and sustainable, we introduce the MSR 2025 Blended PC model. 
This model builds on the success and insights from similar initiatives outside the field (e.g., ICML) and the Shadow and Junior PC initiative at MSR. 
This new model for the Technical Track assigns each paper a core review team of two regular PC members and one junior PC member
To assess the perceived quality of reviews, we survey authors of both accepted and rejected papers, complementing ICML's evaluation using indirect indicators (e.g., review length, discussion activity) and meta-reviewer assessment~\cite{stelmakh2021novice} 

Our findings show that authors often rated junior PC reviews more favorably, despite paper rejections, and could not identify the junior reviewer. This highlights the model’s potential for fair and inclusive participation while still being lightweight. 
This approach continues to inherit the benefits of previous similar initiatives, including: 

\begin{itemize}
\item See examples of actual reviews for the same papers, written by the regular PC members, gaining more experience as a reviewer and learning from the senior researchers how to write better reviews. 
\item Participate in the real-world review process and contribute to the decision-making on papers, thus improving the motivation and engagement of the junior reviewers throughout the process.
\item Get to know how a review cycle is run and how a PC operates.
\item Receive mentorship from experienced PC members.
\item Gain experience in reviewing papers and in understanding the challenges faced by reviewers reading multiple papers, which may not always be in their area of expertise.
\item Submitting high-quality reviews makes a junior reviewer a more likely candidate for future PCs of the technical track of MSR (and/or other software engineering conferences) or otherwise.
\item See both strong and weak papers at the submission stage.
\item Have a chance to read top-notch papers in your area of expertise before they are published.
\end{itemize}

\section{Blended PC Review Model} \label{sec:model}

\textbf{Motivation.} Different from the MSR Junior PC review processes in 2023 and 2024, which involved three regular PC members and two junior PC members 
MSR 2025 has shifted to a Blended PC model of two regular and one junior PC member. This brings the core review team of a given paper back to three, in line with the MSR review process before 2021 and with other conferences. We anticipate the change towards $2$+$1$ to provide substantial benefits to authors, Junior PC members, and regular PC members, driven by the following reasons:

\begin{itemize}
\item \textbf{Reducing Review Workload.} The previous model, requiring five PC members per paper, placed a substantial review burden on both junior and regular PC members, potentially compromising the quality of the reviews due to the tight 1-month review schedule of MSR. By reducing the number of PC members from $3$+$2$ to $2$+$1$, the review workload for each PC will be lowered, thereby expecting to enhance the quality of the reviews.
\item \textbf{Improving Authors' Responses.} MSR provides an author response period to allow authors to clarify their submissions. Previously, with five reviewers per paper, authors struggled to produce high-quality responses within the 750-word limit. We expected that the reduction to three reviews (two regular, one junior) would likely help authors craft more effective and manageable responses within the word limit.
\item \textbf{Improving Review Quality} Based on our experience with the MSR Shadow PC in 2021 and 2022 and MSR Junior PC in 2023 and 2024, we expected that the addition of a junior PC to each paper would increase the overall quality of reviews that the authors receive, since junior reviewers typically have a deep understanding of recent topics, and can thus provide in-depth technical feedback on the subject.
\end{itemize}
 
The Junior PC co-chairs play an essential role in ensuring that the quality of the review feedback is not compromised, using a combination of practices. 
For MSR 2025, Junior PC members received training from senior researchers and/or past MSR distinguished reviewers on how to write and provide constructive reviews for software engineering papers, what to look for, what to avoid, and the suggested review structure aligned with the MSR review criteria. 
They also received guidance from the MSR PC co-chairs and Junior PC co-chairs on the expected review quality, confidentiality, and ethics standards, how to write good reviews, and how to participate in discussions (see 
ACM reviewers' responsibilities\footnote{\url{https://www.acm.org/publications/policies/peer-review}}. 
Finally, Junior PC members also received feedback on how to improve their reviews before the rebuttal notification, from the discussion lead, who is selected from two technical track regular PC members and also acts as a mentor. 
The discussion lead performs quality assurance to ensure that the reviews are of high quality and constructive. 

\textbf{Review Details for MSR 2025.} As all submissions to the MSR research track are reviewed jointly by both regular and junior PC members, as part of the same process, the Junior PC 
fully engages in the review process, including reviewing author responses, participating in PC discussions, and reaching a decision on the paper.
Most of the papers (110 out of 157 papers, ~70\%) received two reviews from two regular PC members and one review from a Junior PC member, with the remaining papers having three regular PC members. We recruited 110 junior PC members, and hence the distribution.
This represents a significant evolution in the role of the Junior PC, with their evaluations contributing substantially more to the decision-making process compared to previous models.
The final decision is made by consensus among all reviewers. 

\textbf{Junior PC Member Responsibilities.}
Junior PC members are requested to commit themselves to writing their own detailed and rigorous reviews for papers assigned to them by the allotted deadline. This timely review commitment is essential to the well-functioning of the Blended PC. Candidates who might be unable to fulfil their reviewing duties should refrain from applying. 

This year, each Junior PC member was expected to review one paper, submit on-time reviews, and actively participate in the online discussions of their assigned papers. Junior PC members must follow the ethical standards of peer review, respect the anonymity of the review process, not share which papers they have reviewed or solicit sub-reviews, and must not use Generative AI (new for 2025). 

\textbf{Blended PC Eligibility.} The Blended PC is open to PhD students, post-docs, new faculty members, and industry practitioners experienced in software engineering or mining software repositories research, especially those who have not yet served in a program committee of the technical research track (or the main track) of the international SE conferences (e.g., ICSE, FSE, ASE, MSR, ICSME, SANER). We selected the Junior PC members based on their research experience, and also strived to ensure diversity and inclusion (see Tables \ref{tab:expertisebytopic},~\ref{tab:expertisebydatasource}, and~\ref{tab:expertisebymethods} in the Appendix)..

\textbf{Junior PC Member Recruitment.} We used self-nomination as a recruitment tool for the Junior PC members. We ran a one-month recruitment campaign (October 18th, 2024 -- November 20th, 2024, AoE) through various social media channels, e.g., SEWORLD, X (formerly Twitter), and Facebook. 
After collecting self-nominations, we sent out formal invitations to join the Blended PC on December 1st, 2024. We were overwhelmed with the positive reaction from the MSR community, with a total of 248 self-nominations, which is 26\% higher than MSR 2023's 197 Junior PC applications, 123\% higher than MSR 2022's 111 Shadow PC applications, and 53\% higher than MSR 2021's 162 Shadow PC applications. 

After carefully reviewing each nomination, we selected 110 applicants as MSR 2025 Junior PC members (an acceptance rate of 44\%), giving priority to applicants in the later stages of their PhD program and ensuring that we put together a diverse Junior PC that reflects the geographic, demographic, and research area diversity of the MSR community. Below are the summary statistics of the MSR 2025 Junior PC members:
\begin{itemize}
\item Acceptance Rate: 110 Junior PC members out of 248 applications (44\% acceptance rate) are accepted.
\item Gender Diversity: 22\% are women/non-binary, which is similar to the gender ratio of the total applications (24\%).
\item Occupation Diversity: 72\% are PhD students, 11\% are postdocs, 15\% are junior faculty members, and 2\% are industry practitioners.
\item Expertise Diversity: Our Junior PC members cover the entire range of topics, data sources, and research methods relevant to the MSR 2025 Call for Papers~\footnote{\url{https://2025.msrconf.org/track/msr-2025-technical-papers?#Call-for-Papers}}, with the selection ratio ensuring a fair and balanced distribution of the expertise across all research topics (see Tables \ref{tab:expertisebytopic},~\ref{tab:expertisebydatasource}, and~\ref{tab:expertisebymethods} in the Appendix).
\end{itemize}

\textbf{Junior PC Member Training.} Prior to the MSR submission deadline, all PC members, including the junior reviewers, have received guidance on review quality, confidentiality, and ethics standards, how to write good reviews, and how to participate in discussions (refer to the ACM peer review policy\footnote{\url{https://www.acm.org/publications/policies/peer-review}}). 
For 2025, we also introduced a Junior PC member training session. 
The Junior PC co-chairs invited Professor Alexander Serebrenik to share their experiences and tips on how to write a good review, with some anti-patterns to avoid.
The video recording of this training session is available on YouTube\footnote{\url{https://www.youtube.com/watch?v=B5SktafFSiY}}.
The training session was specially designed for PhDs, postdocs, and other early-career researchers and practitioners, but was also open to anyone interested in strengthening their review skills. 
In addition to this live virtual training session, the Junior PC members received review guidelines that outlined the details on the review process, criteria, timelines, expectations, and more from the MSR PC co-chairs.

\section{Survey Design} \label{sec:survey}


With the newly introduced Blended PC model of 2 regular PC + 1 Junior PC members, it is important to understand how the authors perceive this change of review model as well as its potential impact on review quality.
To address this gap, we conducted a qualitative study to answer the following two \emph{research questions}: 

\begin{itemize}
    \item \textbf{RQ1: Which review model is preferred by the authors?}
    \item \textbf{RQ2: Does the review quality of Junior PC members differ from that of regular PC members?}
\end{itemize}

To answer the research questions, we created and distributed anonymous, voluntary surveys to the authors of papers submitted to the MSR 2025 Technical Track.
The survey is designed as a cross-sectional study where participants provided their responses at one fixed point in time, and we encouraged one author per paper to complete the survey. 
The survey aimed to collect feedback on the perceived review quality, on the opinions regarding which review model best fits the MSR Technical Track, and the ability to distinguish between reviews provided by junior and regular PC members. 
As such, the survey consisted of three closed-ended questions and three open-ended questions.

For the closed-ended questions, we used multiple-choice questions and a 5-point Likert scale.
\begin{itemize}
    \item (Q1) Which of the following two review models would you prefer for the MSR Technical Track?
    \item (Q2) How would you rate the (overall) review quality (for Reviewer A/B/C)?
    \item (Q3) Could you guess the junior PC on your paper?
\end{itemize}

Each of these questions was followed by an open-ended question for the rationale, allowing the participants to leave additional free-text comments to elaborate more on their answers and provide further feedback to inform future decisions. 
We used Google Forms to conduct our survey in an online setting.
The survey took approximately 5 minutes to complete and was completely anonymous.
To verify the completeness of the responses to our survey (i.e., whether all questions were appropriately answered), we manually reviewed all of the open-ended questions. 

We contacted 426 unique authors from 161 papers, soliciting one response per paper. In the end, we received 73 responses, which cover 65 papers at a 40\% response rate. For some papers, we received two responses, which we kept for diversity of perspectives. Our responses represent insights from 46 rejected and 20 accepted papers. 43 of these 65 papers had one junior PC member each, while the remaining 22 papers had all three regular PC members.


\section{Survey Results} \label{sec:results}


\subsection{RQ1: Which review model is preferred by the authors?}

\begin{figure}[t]
\includegraphics[width=\columnwidth]{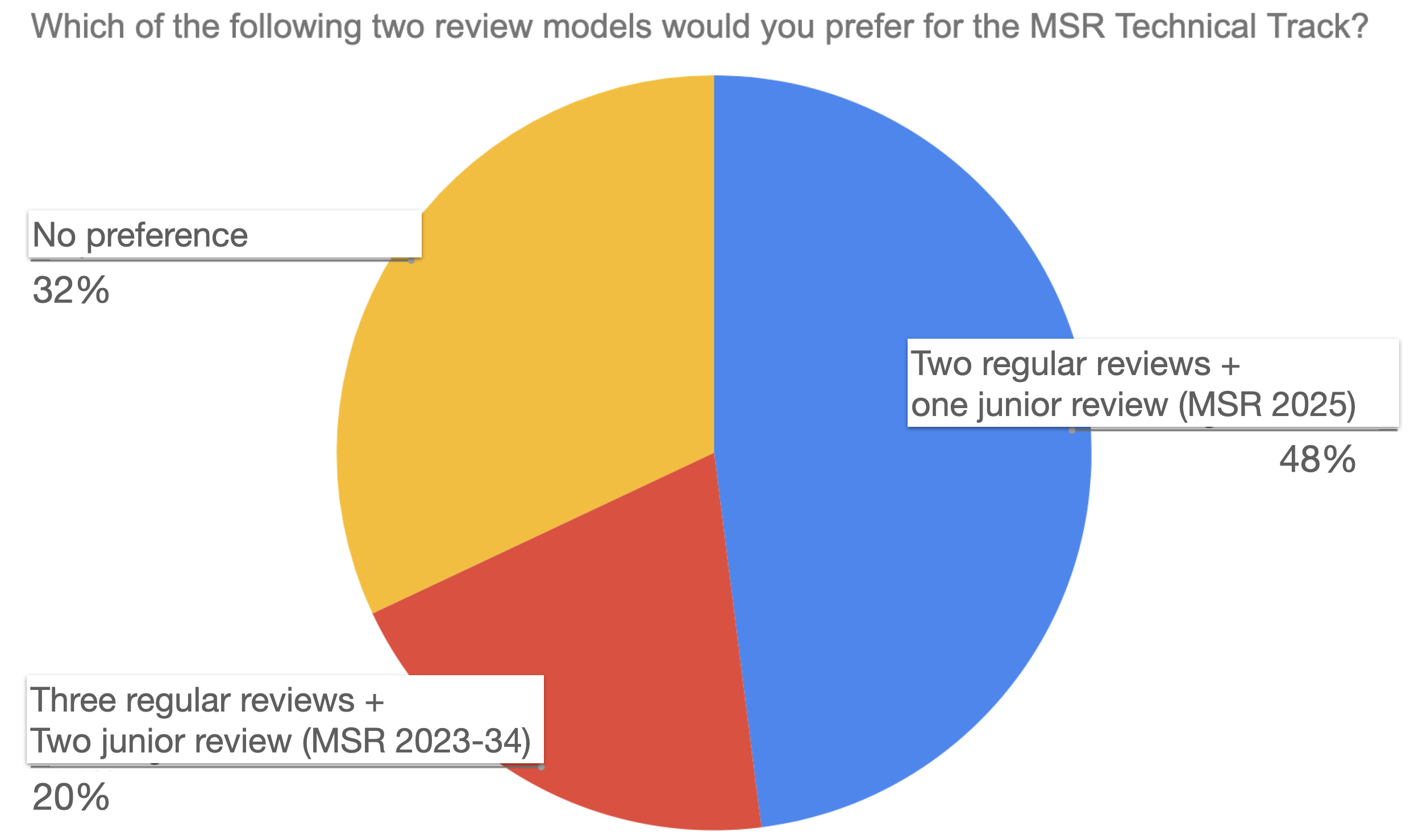}
\caption{Preference regarding the review models to adopt for the MSR Technical Track. }
\label{fig:prefReviewModel}
\end{figure}

When asked which review model best fits the MSR technical track review process, 48\% of authors suggest adopting the new Blended model with two regular PC members and one junior member as core review team (see Figure \ref{fig:prefReviewModel}). 
Furthermore, 32\% expressed no preference, while 20\% declared they would like to see the reintroduction of the prior 3+2 model adopted in MSR 2023 and 2024, where the two Junior PC members were not part of the core review team of a given paper. 

Respondents indicating a preference for the Blended model 
generally support having three reviewers per paper, as this creates an odd number that helps avoid deadlocks in decision-making and provides sufficient coverage without overwhelming authors with too many opinions to reconcile. 
Regarding junior reviewers, the responses showed mixed opinions and presented different perspectives. Several respondents strongly favor the inclusion of junior reviewers, as this was seen as a valuable professional training practice. One respondent specifically mentioned the benefit of valuing the junior PC members' opinions, with the possibility of them playing a decisive role in case of disagreements between the seniors.

One respondent reported a moderate concern that junior reviewers might feel pressure to be overly critical to establish their credibility, potentially leading to harsh rejections. However, the Blended PC review model 
is seen as capable of mitigating any risks deriving from less experienced reviewers taking part in the process, with a balanced approach ensuring the inclusion of diverse perspectives while maintaining an efficient review process.  

As for the supporters of the prior 3+2 review model, some of the open-ended responses highlight the trade-off between maintaining the usual high MSR review quality standards and including junior community members in what is perceived as a valuable training program. Some of the respondents suggest increasing the number of reviewers to reduce randomness in review outcomes and provide more balanced input, although this would increase review load in the community. 

Finally, authors expressing no preference between the two models also provided some suggestions for alternative models. One of the respondents suggests a new possible option comprising three regular and one junior PC member. While increasing the overall review load, such a model may mitigate threats due to the junior reviewer not having enough influence in the discussion, thus following the more experienced reviewers, who would end up dominating the discussion.

\subsection{RQ2: Is there a difference between the review quality of Junior and Regular PC Members?}

\subsubsection{Perceived Review Quality}
\begin{figure}[t]
\includegraphics[width=\columnwidth]{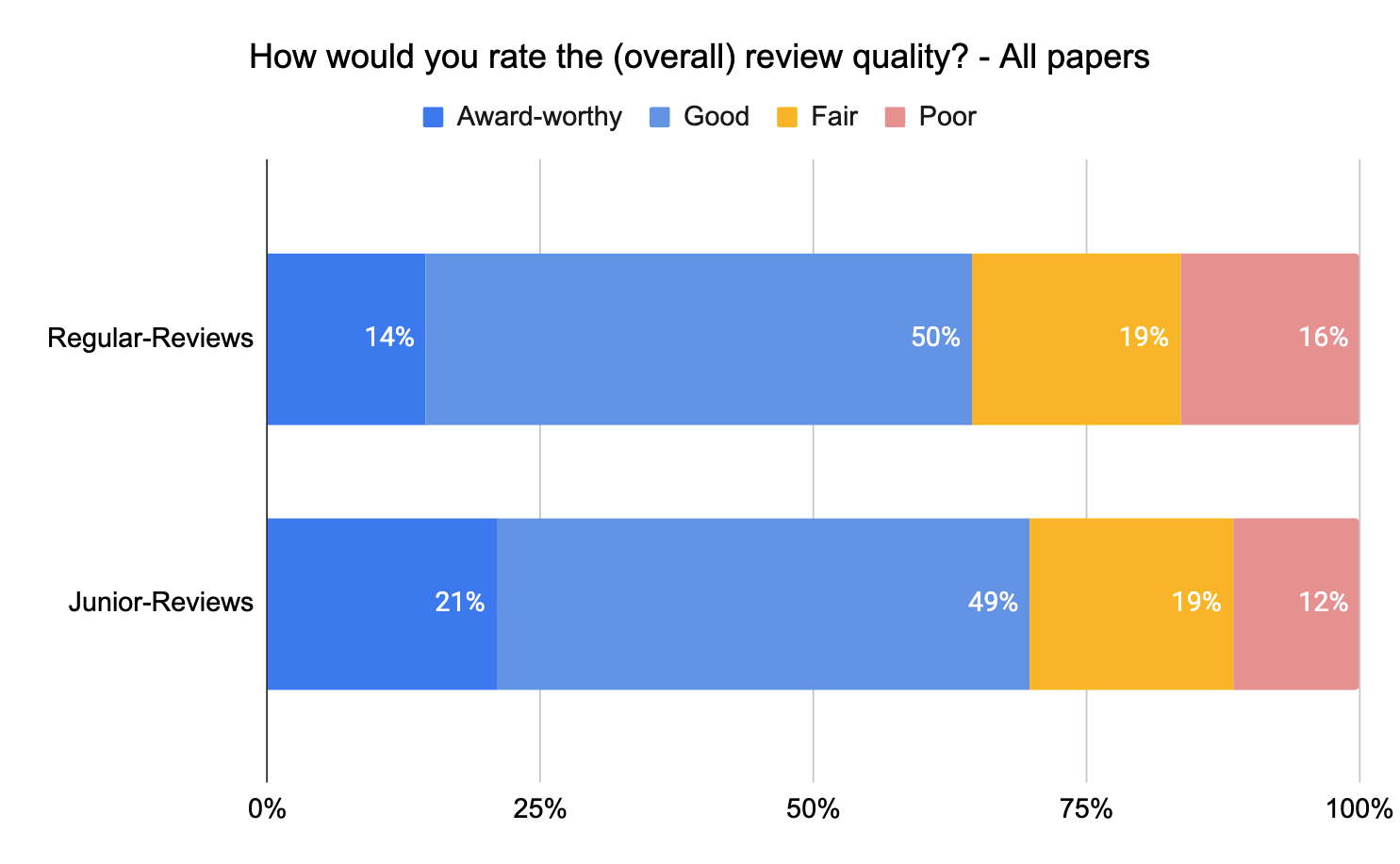}
\caption{Authors' evaluation of the overall review quality.}
\label{fig:reviewQuality}
\end{figure}

Authors were requested to provide an overall evaluation of the quality of each of the three reviews received for their papers. The authors were not aware of which review, if any, was submitted by the Junior PC member; only we had this mapping.
In Figure \ref{fig:reviewQuality}, we report a bar chart illustrating both the absolute numbers and the proportional differences between the two review categories. 

The results show that the authors were generally satisfied with the quality of reviews provided by both regular PC and junior PC, with the majority of reviews being evaluated as either ``Good" (in light blue) or ``Award-worthy"  (in dark blue) quality, thus confirming that the overall perceived quality of the reviews is at least as good (if not better) as the high standard expectations of the authors submitting to the MSR Technical Track. 

In particular, 14\% regular reviews were rated as ``Award-worthy", while 50\% were rated as ``Good". As for junior reviews, we observe a slightly better distribution with 21\% reviews rated as ``Award-worthy"  and 49\% rated as ``Good". 
Interestingly, while junior reviews have fewer total ratings (43 junior reviews were evaluated, compared to 152 for regular reviews), they show a higher percentage of top-quality ratings, with an overall 70\% of respondents rating their reviews as either ``Award-worthy" or ``Good", with only 12\% rating them as ``Poor." What further strengthens these results is that the majority of survey respondents were authors of rejected papers (46 authors of rejected papers vs. 20 authors of accepted papers, 
yet most considered the Junior PC member reviews to be of ``Award-worthy" and ``Good" quality.

This finding suggests that junior reviews' overall quality and commitment hold potential for equaling the highest quality standard, despite them being involved for the first time in the MSR review process (same as in the MSR 2021 Shadow PC~\cite{thongtanunam2021shadow}).
This finding becomes even more significant considering our committee composition --- 110 junior PC members alongside 135 regular PC members. The high-performance metrics from Junior PC members, despite their numerical minority, suggest that their contribution was crucial and an enrichment in the review process. 

This finding is further corroborated by the fact that the reviews provided by junior PC members appear not to be identifiable. When asked to guess which review was provided by the junior PC member, only 16\% of responders could guess the correct answer. Furthermore, we did not observe any substantial differences in the average review scores provided by the two groups, with regular and junior PC members providing, on average, an overall review score of 2.26 and 2.18 (on a scale of 1 to 5), respectively. 

\subsubsection{Perceived Review Quality of Accepted versus Rejected Papers}

To investigate whether experiences with reviews differ based on paper acceptance status, we conducted a follow-up analysis comparing perspectives from authors of accepted versus rejected papers. Note that since the survey was administered during the author response period, participants were unaware of final decisions at the moment of responding to the survey. However, the reviewers' inclinations regarding likely outcomes could often be inferred, so we decided to break down the analysis of feedback for accepted vs. rejected papers to check if the authors' perception might have influenced their survey.

\begin{figure}[t]
\includegraphics[width=\columnwidth]{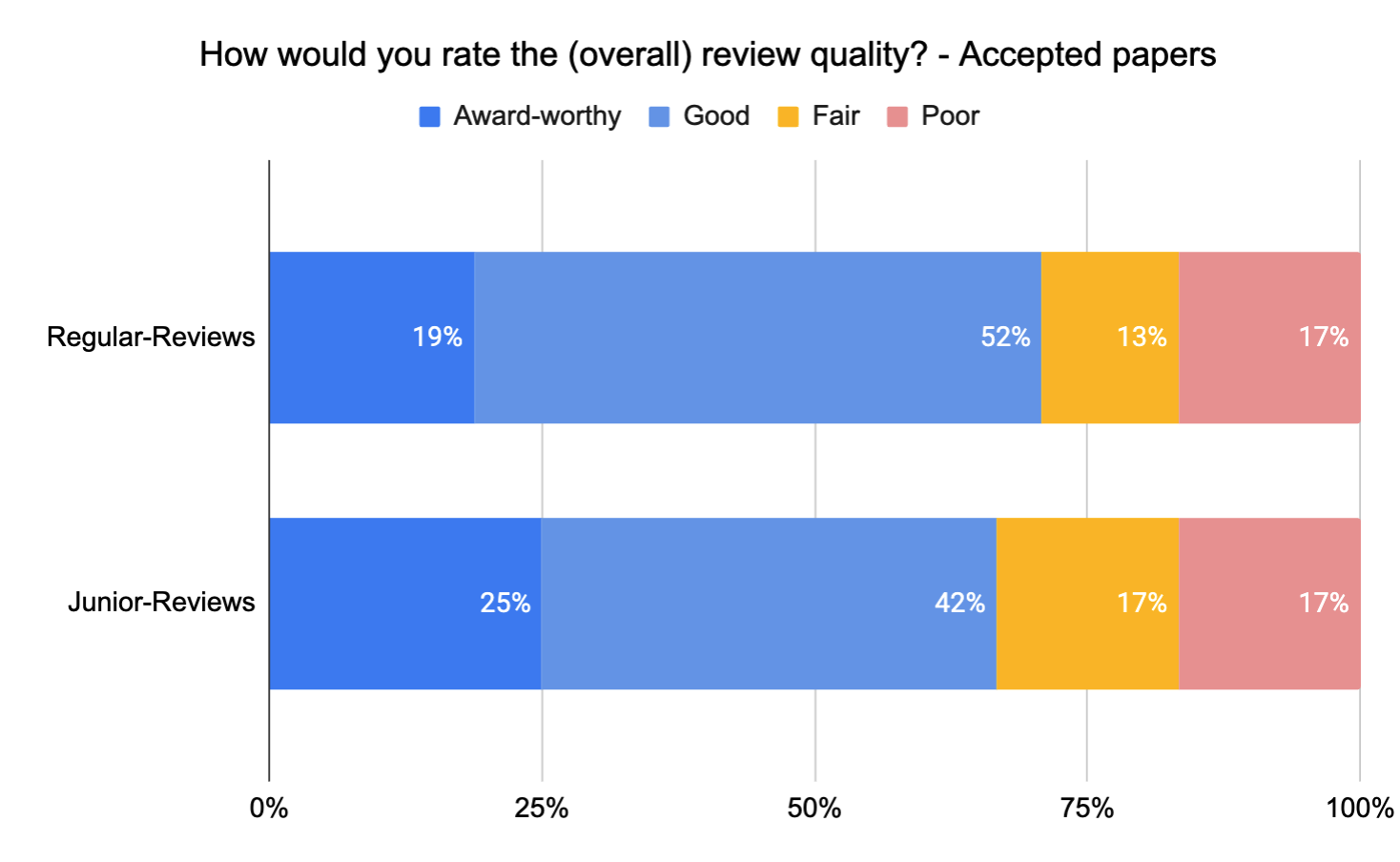}
\caption{Authors' evaluation of the overall review quality for accepted papers.}
\label{fig:reviewQuality:accepted}
\end{figure}

\begin{figure}[t]
\includegraphics[width=\columnwidth]{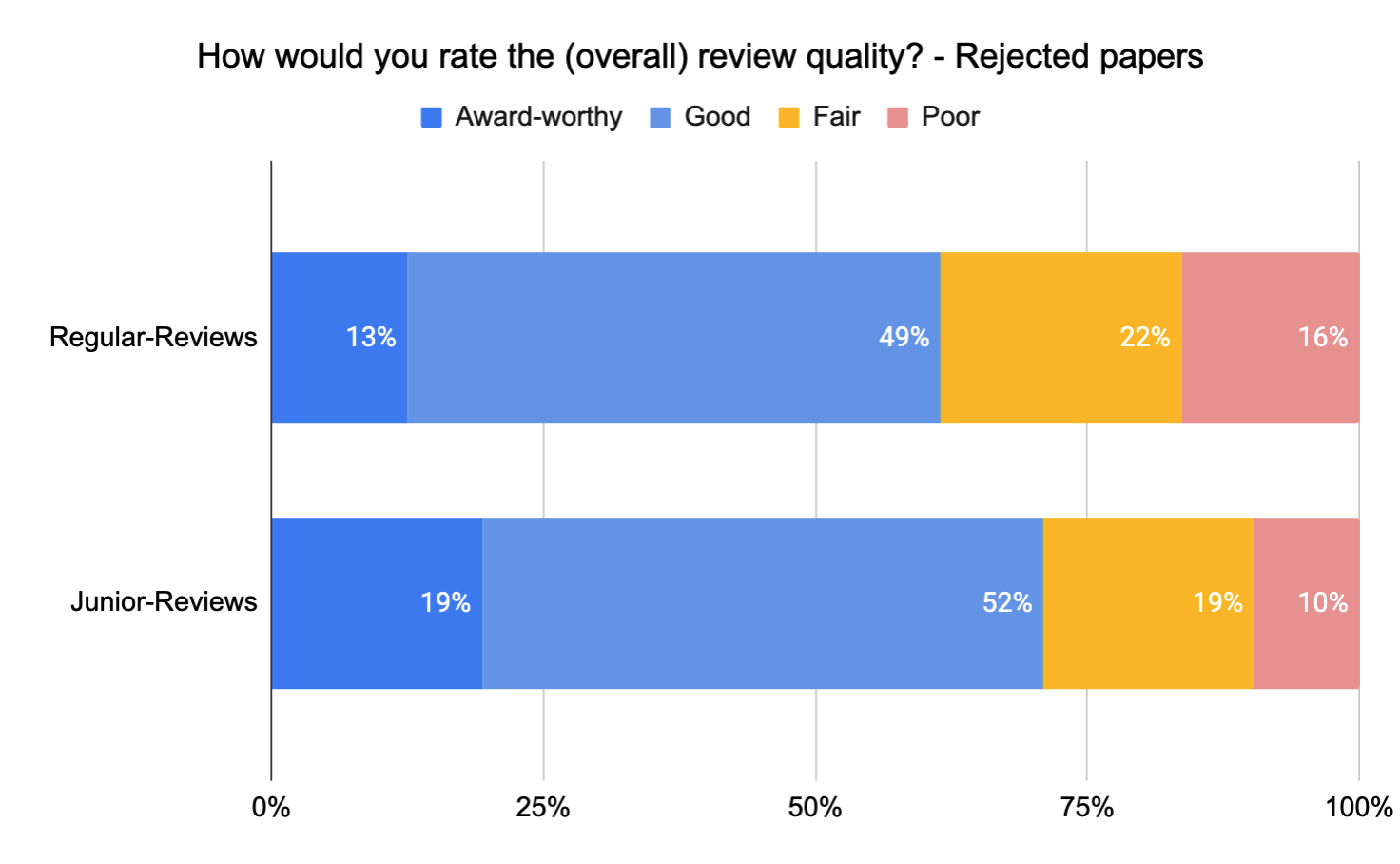}
\caption{Authors' evaluation of the overall review quality for rejected papers.}
\label{fig:reviewQuality:rejected}
\end{figure}

In Figure~\ref{fig:reviewQuality:accepted} and Figure~\ref{fig:reviewQuality:rejected}, we report the results of the analysis for accepted (including conditional acceptance) and rejected papers, respectively. 
When comparing the overall distribution (see Figure \ref{fig:reviewQuality}) to accepted papers (see Figure \ref{fig:reviewQuality:accepted}), the results reveal a generally more favorable perception of the review quality among authors whose papers were ultimately accepted. For regular reviewers, authors of accepted papers demonstrated increased satisfaction, with award-worthy ratings rising from 14\% to 19\% and good ratings improving marginally from 50\% to 52\%. At the same time, fair ratings decreased from 19\% to 13\%, while poor ratings remained comparable (16\% vs. 17\%).

The pattern for junior reviews reveals a slightly different picture, with authors of accepted papers rating junior reviews as award-worthy at a higher rate (25\% vs. 21\% overall), though good ratings declined from 49\% to 42\%. Surprisingly, the poor rating category increased from 12\% to 17\%, in spite of the positive outcome for the paper. 

Interestingly, for junior reviews of rejected papers, the data reveals a counterintuitive pattern. In fact, authors of rejected papers provided more favorable ratings in positive categories, with good ratings increasing from 49\% (see Figure \ref{fig:reviewQuality}) to 52\% (see Figure \ref{fig:reviewQuality:rejected}) 
and poor ratings decreasing from 12\% to 10\%. Even if award-worthy ratings slightly declined from 21\% to 19\%, the overall amount of positive reviews remains comparable (70\% for all reviews and 71\% for rejected papers). 

This evidence is not confirmed for regular reviews, for which authors of rejected papers showed slightly lower satisfaction, with award-worthy ratings declining from 14\% to 13\% and fair ratings increasing from 19\% to 22\%. Good ratings remained remarkably stable at approximately 49-50\%, while poor ratings showed minimal variation. 

These findings seem to suggest that Junior PC members probably invested more effort in providing good-quality reviews for papers that were eventually rejected compared to accepted ones. 
Furthermore, these findings suggest that authors' perceptions of review quality may indeed be influenced by their anticipation of acceptance outcomes, even if mixed findings are observed for regular and junior PC members. The generally more positive assessment of review quality among authors of accepted papers aligns with the expectation that perceived review inclinations could have influenced survey responses. However, the nuanced patterns and, in particular, the mixed results for the rejected papers, indicate that the relationship between acceptance anticipation and review quality perception is more complex than a simple positive correlation.


\subsubsection{Characteristics of `poor' and `fair
' reviews}
In the following, we report the results of a qualitative analysis of the open-ended answers in which the survey respondents reported the reasons for providing either a 'fair' or 'poor' evaluation of what they perceived as lower-quality reviews. To this aim, we analysed 37 open-ended responses, of which nine were associated with Junior PC members' reviews and 28 responses were associated with senior PC members' reviews. We are unable to offer a distinction between junior and regular reviewers, given fewer open-ended responses for junior reviewers. 

Specifically, one of the authors initially proceeded with open coding, thus creating the five codes. Then, another author conducted a second round of coding using the five codes and adding a sixth one. The use of multiple codes was allowed. Overall, a 90\% agreement was observed, and the disagreement on four cases was resolved through discussion. 
The results are reported in Table \ref{tab:codes}, which includes the taxonomy of codes emerging from the qualitative analysis, along with their counts, definition, and example excerpts from the open-ended answers. For six of the items, no justification for the low rating was provided. 

\begin{table*}[]
\begin{tabular}{ | p{0.15\linewidth} |
                   p{0.35\linewidth} |
                   p{0.45 \linewidth} | } 
\textbf{Code} (\textit{count}) &  \textbf{Definition}   & \textbf{Example} \textit{excerpts from the open-ended answers}\\
\hline
\textbf{Lack of domain expertise} (5) &   Reviewers lack sufficient knowledge in the relevant fields & \textit{From the comments, it appeared that Reviewer A did not know about the importance of App review mining and Requirements Engineering, which was an important part on which we based our paper} \\
\textbf{Invalid criticism} (6) &   Criticisms that do not align with accepted research methodologies and with the conference standards or reviewing criteria & \textit{However, in parallel, all (!) reviewers make invalid criticisms according to SIGSOFT Standards (especially B, but also C and to a lesser extent A). It seems they do not know about the scope and inherent limitations of basic research methods.} \\
\textbf{Inattentive review and factual errors} (13) &   Reviewers making factual mistakes or failing to properly examine submitted materials, often resulting in a short and non-informative review & \textit{Furthermore, Reviewer B and C seem to be inattentive reviewers, as they failed to examine the supplementary materials and incorrectly stated that we did not submit a replication package.} \\
\textbf{Lack of constructive feedback} (6) &   Reviewers assign poor ratings without providing helpful suggestions for improvement & \textit{The review lacks any constructive recommendations or suggestions, yet assigns a poor rating, which I find unfair.} \\
\textbf{Inconsistent or unjustified scoring} (3) &   Reviewers assigned or modified scores inconsistently, particularly where there is a misalignment disconnect between the severity of issues reported and the numerical ratings & \textit{The comment from reviewer A ('insignificant contribution') is out of place... If the novelty is difficult to evaluate, how can the contribution be 'insignificant'?} \\
\textbf{LLM use} (2) &   Reviewers provide feedback that appears to be obtained using LLMs for reviewing. & \textit{Reviewer B's comments left many doubts regarding the actual reading/understanding of the paper. For some of the comments, I think an LLMs was used.}
\end{tabular}%
\caption{A taxonomy of codes describing the characteristics of reviews rated as either 'poor' or 'fair'.}
\label{tab:codes}
\end{table*}

As shown in Table \ref{tab:codes}.  The most frequent cause for \emph{poor} and \emph{fair} review ratings is associated with the authors' perception of an inattentive review process, involving also factual errors. In these 13 cases, the respondents reported the reviewers made factual mistakes or failed to properly examine submitted materials (e.g., \textit{``Furthermore, Reviewer B and C seem to be inattentive reviewers, as they failed to examine the supplementary materials and incorrectly stated that we did not submit a replication package."}) 

Invalid criticisms is also reported in six cases, with respondents suggesting that reviewers' criticisms did not align with the accepted research methodologies, probably due to lack of familiarity with the conferences reviewing practices and criteria (e.g., ``\textit{It seems to me that Reviewer C has never reviewed for a conference like MSR.}", or ``\textit{They expect an automated technique to work without any (!) user input (all complain that we offer users the ability to curate and reuse their input files)}").

The unsatisfying review quality is also associated, in six cases, with the lack of constructive feedback. In these cases, the respondents complained about reviewers assigning poor ratings without providing helpful suggestions for improvement (e.g., ``\textit{The review lacks any constructive recommendations or suggestions, yet assigns a poor rating, which I find unfair.}"). 

In some cases the authors also criticized how reviewers assigned scores, particularly in the presence of inconsistencies in the review (e.g., ``\textit{The comment from reviewer A ('insignificant contribution') is out of place... If the novelty is difficult to evaluate, how can the contribution be 'insignificant'?}"), thus ending up with inconsistent or unjustified scoring (three cases).  

In five cases, the respondents reported they believed the reviewers lack the required domain expertise and sufficient knowledge in the relevant fields (i.e., ``\textit{Reviewer A and B do not appear to possess sufficient expertise in LLM and testing.}"). 

Finally, in two cases, the respondents reported suspecting that the reviews were done with the help of LLMs.


\section{Threats to Validity}
We caution readers to consider the following potential threats to the validity of our research while interpreting the results. 

\begin{itemize}
\item While we aimed to collect one survey response per paper, two authors responded for 8 out of 65 papers. We included all responses in our analysis, which may slightly overrepresent some views. However, we do not see this as a concern, as our goal is to capture perceptions of review quality --- something that can legitimately vary even among co-authors of the same paper.

\item We also note that not all reviewers were assessed for the 65 papers analyzed, as respondents could skip any question. However, this limitation should affect accepted and rejected papers equally and thus not bias our comparative analysis.
\end{itemize}

\newpage 

\section{Reflections} \label{sec:reflections}
This section presents our reflections on the review process and outcomes from the perspectives of the Technical Track PC co-chairs and Junior PC co-chairs. We observed the process from various perspectives and, based on this, provide suggestions and recommendations for future editions of the MSR conference and other conferences facing similar challenges, as well as exploring Blended PC as a potential solution. 

Since the adoption of the Junior/Shadow PC program by MSR, several early-career researchers have ``graduated'' and become integral members of the Technical Track's regular PC. For MSR 2025, we invited eligible MSR 2024 Junior PC members who were recognized with Distinguished Reviewer awards to join the MSR 2025 Technical Track PC as regular reviewers. This ensures that we are contributing to the long-term sustainability of the scientific community. 

The relevance of the Junior PC program is also reflected in the lower review loads overall. While junior PC members reviewed one paper on a subject of their expertise, the Technical Track PC reviewed between two and four papers. First, we assigned a paper to junior PC members based on their expertise, followed by assigning papers to the regular PC members.

One space where we felt junior PC (and their training) can improve is in terms of online paper discussion. The two technical track PC co-chairs followed each paper for review quality before rebuttal and the discussion thereafter. In order to reach a fair acceptance decision, the online discussion between reviewers is just as important as writing a high-quality review. While the authors' response suggests that the quality of reviews from Junior PC members was of comparable quality to the technical track PC members, the technical track PC co-chairs felt that some junior PC reviewers could have voiced their opinions during the online discussion more strongly than they did. 

This was a problem for three papers where the other two regular PC members had divergent opinions, and the junior PC member did not express their views due to a lack of confidence in taking sides. Eventually, the situation was resolved by adding an extra (regular) reviewer from the team of rapid response reviewers and offering a conditional acceptance. However, for future editions, our advice is to also train junior PC members on how to participate in discussions and contribute to the decision-making, reducing their fear of intervening. 

Notably, in all three cases of conditional acceptance, the junior PC members served as shepherds, ensuring that the proposed edits in the conditionally accepted papers are appropriate. The choice was inspired by their expertise on the topics and methodologies and their neutral stance in decision-making. 

We were also limited by the HotCRP review platform\footnote{\url{https://hotcrp.com/}} in realizing our vision for the Blended PC review model, with 
to have no distinction between junior PC and regular PC members. 
For example, while we wanted fellow reviewers not to know who is a junior PC, hiding the ‘junior PC’ tag created for management purposes was not possible. Related to this, we could not assign the ‘artifact check’ role to the reviewers with the ‘junior PC’ tag, although we initially planned on distributing the ‘artifact check’ role equally among all PC members. This again has to do with the limitation of the platform, which does not support two tags for one role. 

Another potential ``wishlist'' feature is the ability to survey authors about review quality from within the HotCRP review platform instead of requiring an external platform for this. Our findings support that authors of rejected papers also consider reviews of high quality, if they offer constructive criticism. 

Our final recommendation is that due to the increasing role of Junior PC members in deciding about papers' acceptance, Junior PC co-chairs, similar to the Technical Track PC co-chairs, should not be able to submit any co-authored papers. 
In the previous model, only Technical Track PC co-chairs were not allowed to submit papers to the track. 
We suggest establishing a similar rule 
for the Junior PC co-chairs.

\section{Conclusion} \label{sec:conclusion}
In this paper, we present the Blended PC peer review model implemented at the International Conference on Mining Software Repositories (MSR) 2025, Ottawa, Canada. 
We introduced and evaluated a new Blended PC model (2+1) for the Technical Track, which includes two regular and one junior PC member, which is different from the previous editions' 3+2 model (i.e., three regular and two junior PC members).
To investigate the authors' perception and their review quality, we conducted a survey to answer two research questions: (1) Which review models are most preferred by the authors? and (2) Is there a difference between the review quality of Junior and Regular PC members?

Based on the 73 responses, we found that (1) 48\% 
of authors preferred adopting the new Blended PC (2+1) model versus 20\% for the previous model; (2) the authors perceived the junior PC members' review quality as at least good if not better, without knowing which reviewer was junior PC member
and (3) only 16\% of respondents could correctly guess the review of a Junior PC member. 
The high satisfaction with review quality from both Junior and regular PC members and the low rate of correct identification suggest that junior reviewers delivered reviews indistinguishable in quality from their senior counterparts, supporting the value of involving early-career researchers in the peer review process and warranting future adoption of this model.

\bibliographystyle{ACM-Reference-Format}
\bibliography{biblio}




\begin{table*}
\caption{Expertise of Junior PC members by topics.}
\label{tab:expertisebytopic}
\resizebox{\textwidth}{!}{%
\begin{tabular}{|p{8cm}|ccc|cc|}
\hline
& \multicolumn{3}{c|}{\textbf{Applications}}                                                               & \multicolumn{2}{c|}{\textbf{Acceptance}}                                             \\ \hline
\textbf{By topics}                                     & \multicolumn{1}{c|}{\textbf{Expert (+2)}} & \multicolumn{1}{c|}{\textbf{Familiar (+1)}} & \textbf{TOTAL} & \multicolumn{1}{c|}{\textbf{SELECT}} & \multicolumn{1}{l|}{\textbf{Selection Ratio}} \\ \hline
API design and evolution                               & \multicolumn{1}{c|}{20}                   & \multicolumn{1}{c|}{71}                     & 91             & \multicolumn{1}{c|}{33}              & 36\%                                          \\ \hline
App store analysis                                     & \multicolumn{1}{c|}{25}                   & \multicolumn{1}{c|}{39}                     & 64             & \multicolumn{1}{c|}{21}              & 33\%                                          \\ \hline
Cloud-based and distributed systems                    & \multicolumn{1}{c|}{25}                   & \multicolumn{1}{c|}{42}                     & 67             & \multicolumn{1}{c|}{17}              & 25\%                                          \\ \hline
Debugging, fault localization, and fault prediction    & \multicolumn{1}{c|}{47}                   & \multicolumn{1}{c|}{95}                     & 142            & \multicolumn{1}{c|}{49}              & 35\%                                          \\ \hline
DevOps and release engineering                         & \multicolumn{1}{c|}{26}                   & \multicolumn{1}{c|}{52}                     & 78             & \multicolumn{1}{c|}{22}              & 28\%                                          \\ \hline
Distributed and collaborative software engineering     & \multicolumn{1}{c|}{32}                   & \multicolumn{1}{c|}{58}                     & 90             & \multicolumn{1}{c|}{21}              & 23\%                                          \\ \hline
Energy profiling                                       & \multicolumn{1}{c|}{6}                    & \multicolumn{1}{c|}{17}                     & 23             & \multicolumn{1}{c|}{6}               & 26\%                                          \\ \hline
Ethical concerns (bias, fairness, explainability, etc) & \multicolumn{1}{c|}{27}                   & \multicolumn{1}{c|}{69}                     & 96             & \multicolumn{1}{c|}{35}              & 36\%                                          \\ \hline
Human aspects                                          & \multicolumn{1}{c|}{42}                   & \multicolumn{1}{c|}{56}                     & 98             & \multicolumn{1}{c|}{29}              & 30\%                                          \\ \hline
Legal aspects (licenses, copyright, etc)               & \multicolumn{1}{c|}{11}                   & \multicolumn{1}{c|}{26}                     & 37             & \multicolumn{1}{c|}{10}              & 27\%                                          \\ \hline
Machine learning for software engineering              & \multicolumn{1}{c|}{107}                  & \multicolumn{1}{c|}{85}                     & 192            & \multicolumn{1}{c|}{35}              & 18\%                                          \\ \hline
Performance analysis and testing                       & \multicolumn{1}{c|}{48}                   & \multicolumn{1}{c|}{79}                     & 127            & \multicolumn{1}{c|}{38}              & 30\%                                          \\ \hline
Privacy and security                                   & \multicolumn{1}{c|}{46}                   & \multicolumn{1}{c|}{62}                     & 108            & \multicolumn{1}{c|}{28}              & 26\%                                          \\ \hline
Program comprehension                                  & \multicolumn{1}{c|}{32}                   & \multicolumn{1}{c|}{70}                     & 102            & \multicolumn{1}{c|}{37}              & 36\%                                          \\ \hline
Recommender systems                                    & \multicolumn{1}{c|}{27}                   & \multicolumn{1}{c|}{59}                     & 86             & \multicolumn{1}{c|}{32}              & 37\%                                          \\ \hline
Software development processes                         & \multicolumn{1}{c|}{53}                   & \multicolumn{1}{c|}{77}                     & 130            & \multicolumn{1}{c|}{32}              & 25\%                                          \\ \hline
Software ecosystems                                    & \multicolumn{1}{c|}{38}                   & \multicolumn{1}{c|}{63}                     & 101            & \multicolumn{1}{c|}{33}              & 33\%                                          \\ \hline
Software engineering for machine learning              & \multicolumn{1}{c|}{75}                   & \multicolumn{1}{c|}{78}                     & 153            & \multicolumn{1}{c|}{33}              & 22\%                                          \\ \hline
Software maintenance and evolution                     & \multicolumn{1}{c|}{48}                   & \multicolumn{1}{c|}{105}                    & 153            & \multicolumn{1}{c|}{47}              & 31\%                                          \\ \hline
Software measurement and analytics                     & \multicolumn{1}{c|}{33}                   & \multicolumn{1}{c|}{88}                     & 121            & \multicolumn{1}{c|}{43}              & 36\%                                          \\ \hline
Software quality assurance                             & \multicolumn{1}{c|}{52}                   & \multicolumn{1}{c|}{75}                     & 127            & \multicolumn{1}{c|}{38}              & 30\%                                          \\ \hline
Text analysis                                          & \multicolumn{1}{c|}{41}                   & \multicolumn{1}{c|}{76}                     & 117            & \multicolumn{1}{c|}{38}              & 32\%                                          \\ \hline
\end{tabular}}

\caption{Expertise of Junior PC members by data source.}
\label{tab:expertisebydatasource}
\resizebox{\textwidth}{!}{%
\begin{tabular}{|p{8cm}|ccc|cc|}
\hline
& \multicolumn{3}{c|}{\textbf{Applications}}                                                               & \multicolumn{2}{c|}{\textbf{Acceptance}}                                             \\ \hline
\textbf{By repo}                                                    & \multicolumn{1}{c|}{\textbf{Expert (+2)}} & \multicolumn{1}{c|}{\textbf{Familiar (+1)}} & \textbf{TOTAL} & \multicolumn{1}{c|}{\textbf{SELECT}} & \multicolumn{1}{l|}{\textbf{Selection Ratio}} \\ \hline
App stores                                                          & \multicolumn{1}{c|}{30}                   & \multicolumn{1}{c|}{45}                     & 75             & \multicolumn{1}{c|}{23}              & 31\%                                          \\ \hline
Communication data (mailing lists, Gitter/Slack, etc.) & \multicolumn{1}{c|}{25}                   & \multicolumn{1}{c|}{67}                     & 92             & \multicolumn{1}{c|}{34}              & 37\%                                          \\ \hline
Build / CI logs                                                     & \multicolumn{1}{c|}{30}                   & \multicolumn{1}{c|}{45}                     & 75             & \multicolumn{1}{c|}{25}              & 33\%                                          \\ \hline
Code reviews                                                        & \multicolumn{1}{c|}{25}                   & \multicolumn{1}{c|}{67}                     & 92             & \multicolumn{1}{c|}{41}              & 45\%                                          \\ \hline
Execution traces and logs                                           & \multicolumn{1}{c|}{30}                   & \multicolumn{1}{c|}{45}                     & 75             & \multicolumn{1}{c|}{25}              & 33\%                                          \\ \hline
Human interaction data                                              & \multicolumn{1}{c|}{25}                   & \multicolumn{1}{c|}{67}                     & 92             & \multicolumn{1}{c|}{41}              & 45\%                                          \\ \hline
Issue trackers / defect repositories                                & \multicolumn{1}{c|}{30}                   & \multicolumn{1}{c|}{45}                     & 75             & \multicolumn{1}{c|}{25}              & 33\%                                          \\ \hline
Package registries / software releases                              & \multicolumn{1}{c|}{25}                   & \multicolumn{1}{c|}{67}                     & 92             & \multicolumn{1}{c|}{41}              & 45\%                                          \\ \hline
Q\&A websites                                                       & \multicolumn{1}{c|}{30}                   & \multicolumn{1}{c|}{45}                     & 75             & \multicolumn{1}{c|}{25}              & 33\%                                          \\ \hline
Source / test code                                                  & \multicolumn{1}{c|}{25}                   & \multicolumn{1}{c|}{67}                     & 92             & \multicolumn{1}{c|}{41}              & 45\%                                          \\ \hline
Version control history logs                                        & \multicolumn{1}{c|}{30}                   & \multicolumn{1}{c|}{45}                     & 75             & \multicolumn{1}{c|}{25}              & 33\%                                          \\ \hline
\end{tabular}}

\caption{Expertise of Junior PC members by research methods.}
\label{tab:expertisebymethods}
\resizebox{\textwidth}{!}{%
\begin{tabular}{|p{8cm}|ccc|cc|}
\hline
& \multicolumn{3}{c|}{\textbf{Applications}}                                                               & \multicolumn{2}{c|}{\textbf{Acceptance}}                                            \\ \hline
\textbf{By research methods}                                                                                       & \multicolumn{1}{c|}{\textbf{Expert (+2)}} & \multicolumn{1}{c|}{\textbf{Familiar (+1)}} & \textbf{Total} & \multicolumn{1}{c|}{\textbf{Total}} & \multicolumn{1}{l|}{\textbf{Selection Ratio}} \\ \hline
Controlled experiments (e.g., A/B tests, lab studies)                                                                  & \multicolumn{1}{c|}{53}                   & \multicolumn{1}{c|}{74}                     & 127            & \multicolumn{1}{c|}{38}             & 30\%                                          \\ \hline 
Qualitative analysis (e.g., thematic, grounded theory)                                                                            & \multicolumn{1}{c|}{84}                   & \multicolumn{1}{c|}{98}                     & 182            & \multicolumn{1}{c|}{46}             & 25\%                                          \\ \hline
Survey design and analysis                                                                                                                 & \multicolumn{1}{c|}{53}                   & \multicolumn{1}{c|}{74}                     & 127            & \multicolumn{1}{c|}{38}             & 30\%                                          \\ \hline
Statistical analysis \& modeling                                                                                                           & \multicolumn{1}{c|}{84}                   & \multicolumn{1}{c|}{98}                     & 182            & \multicolumn{1}{c|}{46}             & 25\%                                          \\ \hline
Classical machine learning (e.g., regression, SVMs, nearest neighbor, decision trees, naive Bayes, and k-means) & \multicolumn{1}{c|}{53}                   & \multicolumn{1}{c|}{74}                     & 127            & \multicolumn{1}{c|}{38}             & 30\%                                          \\ \hline
Convolutional neural networks                                                                                                              & \multicolumn{1}{c|}{84}                   & \multicolumn{1}{c|}{98}                     & 182            & \multicolumn{1}{c|}{46}             & 25\%                                          \\ \hline
Recurrent neural networks                                                                                                                  & \multicolumn{1}{c|}{53}                   & \multicolumn{1}{c|}{74}                     & 127            & \multicolumn{1}{c|}{38}             & 30\%                                          \\ \hline
Transformers                                                                                                                               & \multicolumn{1}{c|}{84}                   & \multicolumn{1}{c|}{98}                     & 182            & \multicolumn{1}{c|}{46}             & 25\%                                          \\ \hline
Reinforcement learning                                                                                                                     & \multicolumn{1}{c|}{53}                   & \multicolumn{1}{c|}{74}                     & 127            & \multicolumn{1}{c|}{38}             & 30\%                                          \\ \hline
\end{tabular}}
\end{table*}

\end{document}